\documentclass[a4paper,10pt,twoside]{cpc-hepnp}

\usepackage{multicol}
\usepackage{graphicx}
\usepackage{booktabs}
\usepackage{amssymb,bm,mathrsfs,bbm,amscd}
\usepackage[tbtags]{amsmath}
\usepackage[T1]{fontenc}
\usepackage{lastpage}
\usepackage{CJK}
\usepackage{multirow}

\begin{document}

\fancyhead[co]{\footnotesize LIU Mei~ et al: Beam test of a one-dimensional position sensitive chamber on synchrotron radiation}

\footnotetext[0]{Received 30 November 2012}

\title{Beam test of a one-dimensional position sensitive chamber on synchrotron radiation\thanks{Supported by National Natural Science Foundation of China (11275224) }}

\author{%
      LIU Mei$^{1,3,4;1)}$\email{lium@ihep.ac.cn}%
\quad DAI Hong-Liang$^{1,3}$
\quad QI Hui-Rong$^{1,3;2)}$\email{qihr@ihep.ac.cn}%
\quad ZHUANG Bao-An$^{1,3}$\\
\quad ZHANG Jian$^{1,3}$
\quad LIU Rong-Guang$^{1,3}$
\quad ZHU Qi-Ming$^{1,3}$\\
\quad OUYANG Qun$^{1,3}$
\quad CHEN Yuan-Bo$^{1,3}$
\quad JIANG Xiao-Shan$^{1,3}$\\
\quad WANG Ya-Jie$^{2,3}$
\quad LIU Peng$^{3}$
\quad CHANG Guang-Cai$^{3}$
}
\maketitle

\address{%
$^1$ State Key Laboratory of Particle Detection and Electronics, Beijing 100049, China\\
$^2$ CAS Key Laboratory for Biological Effects of Nanomaterials and Nanosafety, Beijing 100049, China\\
$^3$ Institute of High Energy Physics, CAS, Beijing 100049, China\\
$^4$ University of Chinese Academy of sciences, Beijing 100049, China\\
}

\begin{abstract}
One-dimensional single-wire chamber was developed to provide high position resolution for powder diffraction experiments with synchrotron radiation. A diffraction test using the sample of SiO$_2$ has been accomplished at 1W2B laboratory of Beijing Synchrotron Radiation Source. The data of beam test were analyzed and some diffraction angles were obtained. The experimental results were in good agreement with standard data from ICDD powder diffraction file. The precision of diffraction angles was 1\% to 4.7\%. Most of relative errors between measured values of diffraction angles and existing data were less than 1\%. As for the detector, the best position resolution in the test was 138~$\mu$m ($\sigma$ value) with an X-ray tube. Finally, discussions of the results were given. The major factor that affected the precision of measurement was deviation from the flat structure of detector. The effect was analyzed and it came to a conclusion that it would be the optimal measurement scheme when the distance between the powder sample and detector was from 400 mm to 600 mm.
\end{abstract}



\begin{multicols}{2}

\section{Introduction}

Since the advent of MWPC by Charpak\cite{lab1}, wire chambers have been widely used in high energy physics experiments. They can, for example, provide high precision measurement when working as part of time expansion chamber system in TWIST experiment\cite{lab2}. In the last few years, wire chambers are frequently applied on neutron detection because of their low-cost, large-area and reliability\cite{lab3}.
X-ray diffraction is an irreplaceable method for powder crystal lattice measurement. Meanwhile, synchrotron light sources provide intense, tunable and highly collimated radiation which covers from microwaves to hard X-rays. The pulsed time structure in synchrotron X-ray makes it suitable to be applied on biological macromolecules measurement, usually with $0.5\sim2~\overset{\circ}{A}$ wavelength\cite{lab4}. Wire chambers had been used for diffracted X-ray detection. In powder X-ray crystallography, a one dimensional curved wire chamber with delay line readout had been constructed and used\cite{lab5}. In our lab, a single-wire chamber with a FPGA chip's data acquisition was developed for powder X-ray diffraction detection on Synchrotron Radiation and X-ray tubes. The design of detector bas been improved for higher performance and the the beam test has been done at 1W2B laboratory of Beijing Synchrotron Radiation Source in June. In the paper, the details of detector design are presented and the discussion of beam test results are given, and the results show that the one-dimensional wire chamber could be used on SR X-ray diffraction experiment of some samples.

\section{Principle of operation and experimental setup}

\subsection{Principle of X-ray Diffraction}
When the parallel X-ray irradiated a SiO$_2$ powder sample, it would produce diffraction at certain angles. The angles are determined by Bragg's Law:

\begin{equation}
2d\sin \theta = \lambda
\end{equation}

In here 2$\theta$ is the angle between the diffracted and incident directions. Due to the isotropy of the powder, the diffracted X-ray will form a series of cones at 2$\theta$ directions. Used a 2D imaging detector, there would be several bright rings on the plane of active detector. 1D wire chamber detector would detect a series of symmetrical points along x axis and obtain the same angles with 2D imaging detector\cite{lab6}.

\subsection{One-dimensional wire chamber detector}
The wire chamber is made up of an active window£¨along with an upper cathode), an anode wire, the readout strips and the data acquisition electronics. The Figure 1 shows that the schematic of the chamber's structure~\ref{detector}. The sensitive area of the detector is $50mm\times8mm$ with the effective detection length of 200 mm. The window is $200\times10$ mm$^2$, and is made of flat Al foil of 55 $\mu$m thick, and the upper cathode is made of the same material. Under the anode wire of 15 $\mu$m gold plated tungsten with 30 grams tension, the 200 gold-plated strips (the width of 0.5 mm, the length of 30 mm and the pitch of 1 mm) are used as readout pads on the print circuit board. The chamber is filled with working gas, which is usually the mixture of 70\% argon and 30\% carbon dioxide, and the gas pressure is maintained at 1 atm. The upper cathode and readout pad are at zero potential, while the anode wire is maintained at high voltage.

When the X-ray photon is been detected, signals are induced on both anode wire and some readout strips. Induced signals are amplified by pre-amplifiers before being discriminated. Discriminated signals will be transferred to FPGA, which handles a fast processing of signals. In particular, TDC in FPGA will convert signals into digital codes. The pulse width of discriminated signals is proportional to Q value in original signals. According to the corresponding wire number, the centroid method will be used on the digital signals. As a result, the position of gravity center of the hit is obtained. Hit positions information of the photon will be recorded at a data file and transferred to computer.(Fig.~\ref{electronics}).

\begin{center}
\includegraphics[width=7cm]{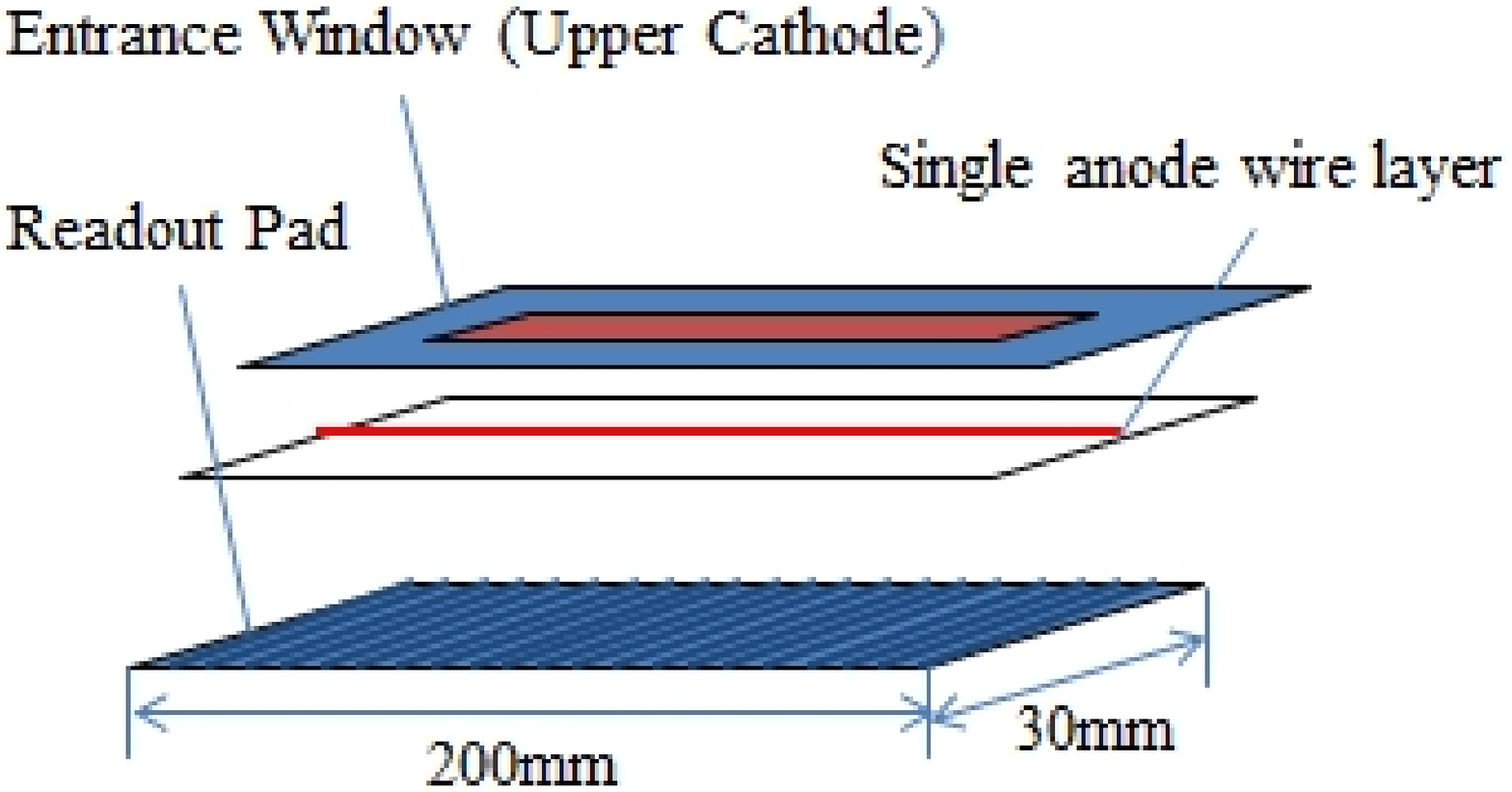}
\figcaption{\label{detector}  A schematic picture showing 1D single-wire chamber structure, the gap between neighboring layers is 4 mm.}
\end{center}

\begin{center}
\includegraphics[width=8cm]{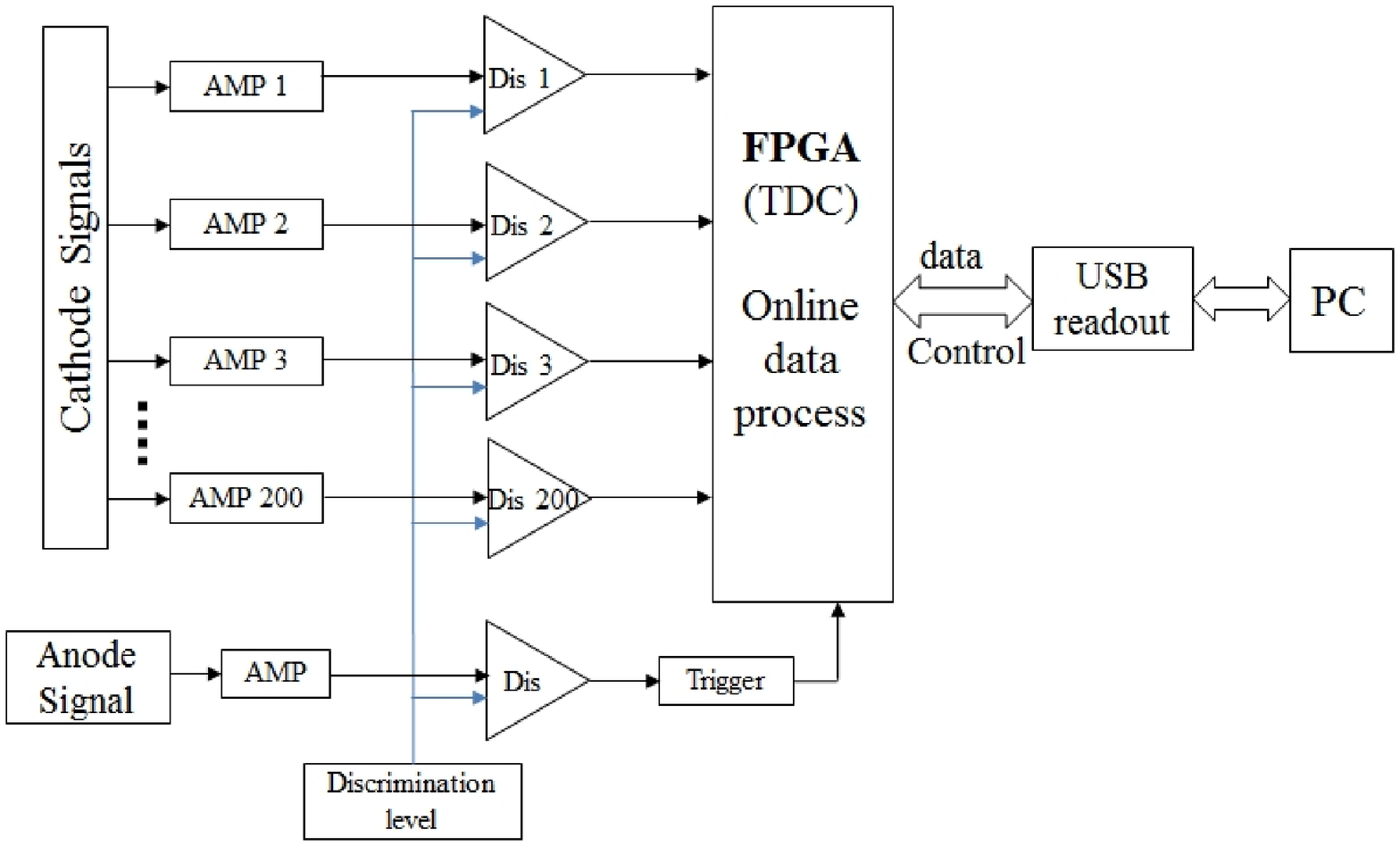}
\figcaption{\label{electronics}  The signal readout electronics of the detector.}
\end{center}

The position resolution of the detector is an important performance for the precision of  diffraction angles. An experiment for its measurement was set up with an X-ray tube. The X-ray collimated through a 20 $\mu$m stainless steel slit, which was placed between the chamber and the X-tube. With 1520 V voltage and 90\%Ar+10\%CO$_2$ gas, the best position resolution measured was 138 $\mu$m in $\sigma$ (Fig.~\ref{Ar-CO2}(a)). The voltage was set according to counting rate curve(Fig.~\ref{Ar-CO2}(b)).

\begin{center}
\includegraphics[width=8cm]{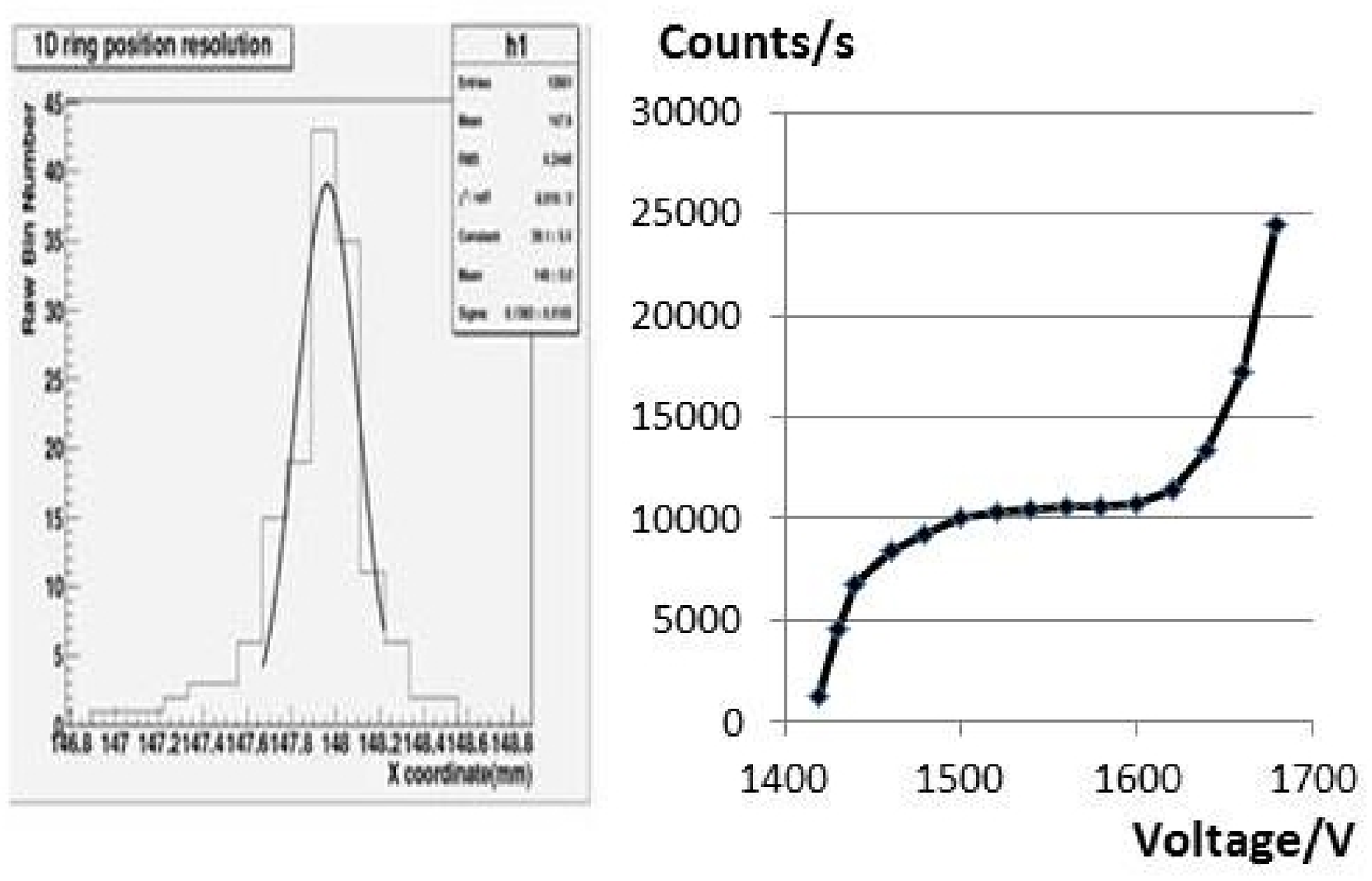}
\figcaption{\label{Ar-CO2}(a) Position resolution with 20 $\mu$m slit @Ar/CO$_2$(90/10). (b) Counting-rate plateau curve of chamber with work gas.}
\end{center}

\subsection{Experimental setup on SR X-ray beam}
At 1W2B laboratory of SR Source, the schematic diagram of diffraction experimental setup on SR X-ray beam is showed in Fig.~\ref{SR}. The center of beam, the 2D collimator, the middle of SiO2 sample and the center of chamber are at the same height with a common X axis position. The lead block is placed as the beam stop for the protection of detector. It is assumed that incident direction is Z and the detector is installed along X. The parallel synchrotron X-ray from a vacuum pipe irradiated the SiO$_2$ sample and produced diffraction. Changed the energy of X-ray from 12 keV to 19 keV and changed the distance between SiO$_2$ sample and detector from 200.5 mm to 232.5 mm. the diffracted X-ray was detected by a single-wire chamber in front of sample. The powder SiO$_2$ used in the test is a standard sample which is NO. 46-1045 in the powder diffraction file-2(pdf-2, from The International Centre for Diffraction Data(ICDD)).

\begin{center}
\includegraphics[width=8cm]{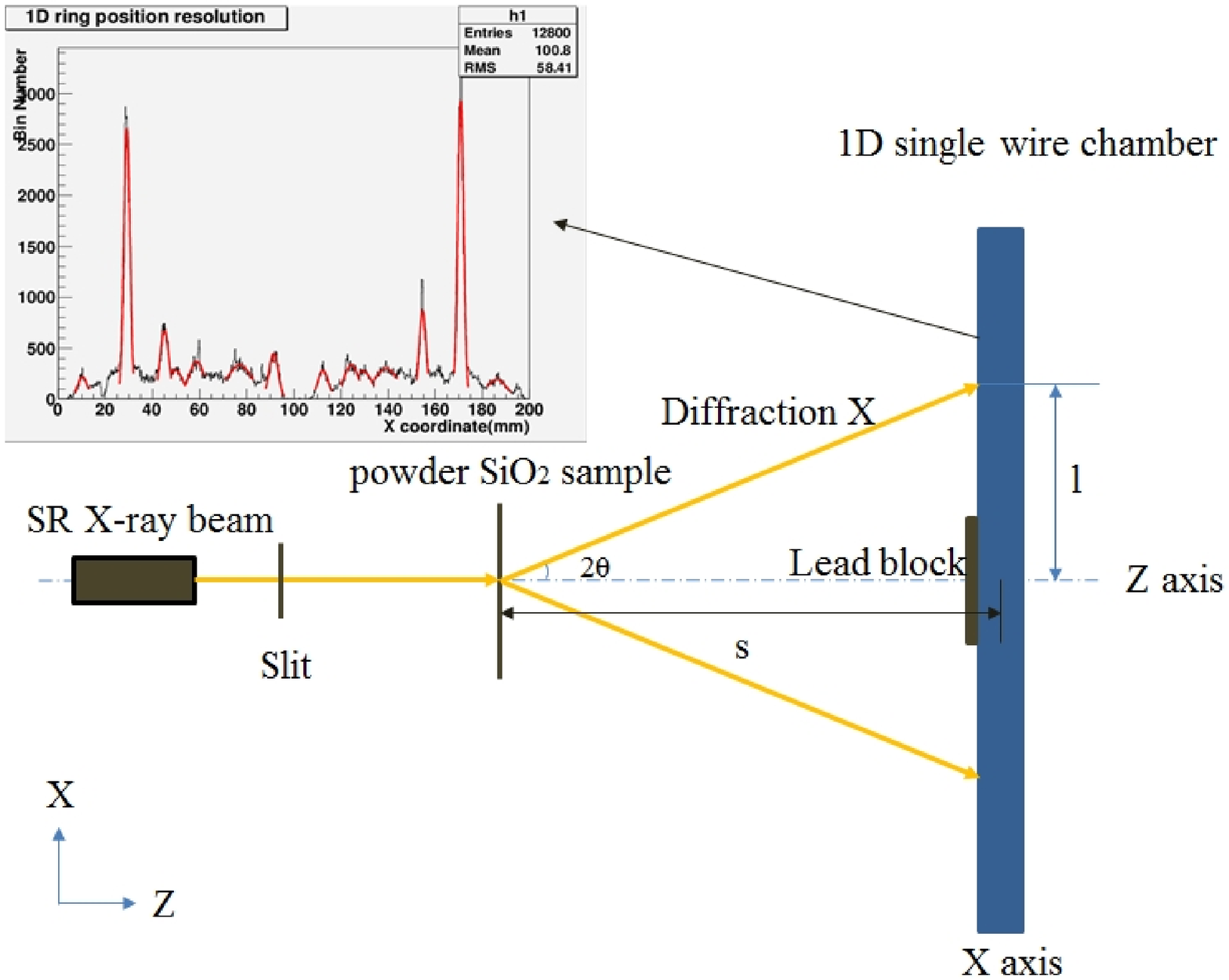}
\figcaption{\label{SR}  A schematic diagram of diffraction experimental setup on SR X-ray beam. The spectrum shows 1D diffraction peaks of sample.}
\end{center}

According to Bragg's Law, the interplanar spacing d is determined by diffraction angle 2$\theta$ with a given wavelength. In that case, the accuracy of measured d depends on the accuracy of measured 2$\theta$. In this test, X-ray wavelength could be known by given energy, then the diffraction angles could be calculated by Bragg's Law.

\section{Results and discussion}

\subsection{Test results}
In the SR beam test, two parameters have been changed: the energy of X-ray ($E$) and the distance between SiO$_2$ sample and the detector ($S$).

When online data acquisition was done, raw data files were obtained. Each of the files contained two columns of data. One column for fixed 12800 data points within 200 mm at equal distance, the other column recorded counts on each data point. The spectrum of data shows the continuous Gaussian distribution peaks. Offline data processing was accomplished by ROOT\cite{lab7} software. Since the diffraction peaks were symmetric along $x$ axis, only the $+x$ data was processed. The diffraction peak position($l$) information was achieved by applying Gaussian Fitting. Diffraction angles were worked out using trigonometric function£º

\begin{equation}
2\theta = \arctan(\frac{l}{S})
\end{equation}

Assuming $S$ is accurate, according to error transfer formula:

\begin{equation}
N = f(x),~~\sigma_N = \frac{\partial f}{\partial x}\cdot\sigma_x
\end{equation}

It can be deduced that:

\begin{equation}
\sigma_{2\theta} = \frac{S\cdot\sigma_l}{S^2 + l^2}
\end{equation}

$\sigma_l$ is the standard deviation of $l$ obtained from Gaussian Fitting.

ICDD has provided standard $d$ values, as well as $2\theta$ values with a given X-ray wavelength. Fig.~\ref{keV} and Fig.~\ref{distance} presented $2\theta$ values coming from the experiment and ICDD pdf-2\cite{lab8}. The differences between them were measurement errors. The error bars($\sigma_{2\theta}$) of the experimental data represented precision of measurement. According to the data, most of the relative errors between measured values and standard values were <1\%. In the test, most of diffraction peaks' precision of measured values is less than 3\%, but the signal-to-noise ratio of detector is not good enough for small signals, the value of several diffraction peaks is close to 4.7\%. This will be improved in the next test. When 2$\theta$ values converted into d values, relative errors were also <1\%.

\subsection{Discussion}
The measurement errors on the diffraction angles have met expectation, what we focused on is improving the precision of measurement. The source of deviation and its correction were analyzed. Deviation could come from experimental method, which can't be eliminated. Fig.~\ref{error} showed two kinds of deviation originating from the structure of detector. In the former, the flat construction of detector caused imaging aberration $r_a$:

\begin{equation}
r_a = h\cdot\tan2\theta,~~h = 8 mm
\end{equation}

as the detector was 8 mm thick in $Z$ direction. In the latter, the width of the detection in $y$ axis caused detection deviation $r_b$:
\begin{equation}
r_b = l - \sqrt{l^2 - (\frac{y}{2})^2},~~l = s\cdot\tan2\theta
\end{equation}

\begin{center}
\includegraphics[width=7cm]{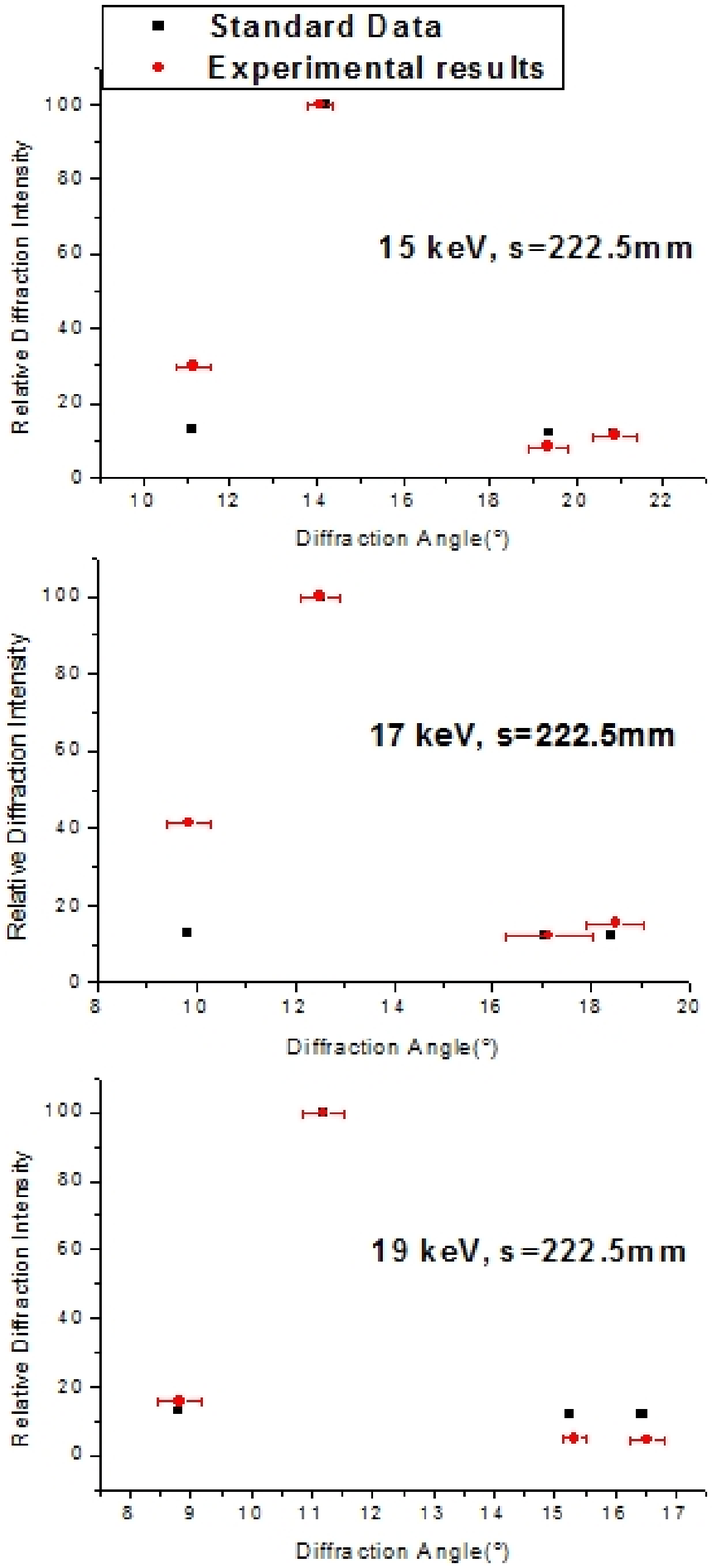}
\figcaption{\label{keV}  Diffraction Angle of sample at various energy(Red points are experimental 2$\theta$, Black squares are pdf-2 standard data).}
\end{center}

where $y$ is the width of the entrance window with 10 mm. It might be noted that $r_a$ and $r_b$ always have the different signs. $r_b$ is confirmed to be negligible. While $r_a$ causes deviation on the position of diffraction peaks, result in damage on 2$\theta$ position resolution.Assuming the hitting events are equally distributed within $r_a$ area, the deviation($\sigma_r$) caused by it is $r_a/\sqrt{12}$. As it was found in the experiment that $2\theta$ precision is enhanced as $2\theta$ increases, in the case of 12 keV X-ray, only considering the minimum diffraction angle. Table.~\ref{mark} shows to what extend $r_a$ affects the final standard deviation on $2\theta$ angles. Although relative $\sigma_{2\theta}$ decreases with increasing $S$, ratio $\sigma_{2\theta}^r/\sigma_{2\theta}$ raises.

\begin{center}
\includegraphics[width=7cm]{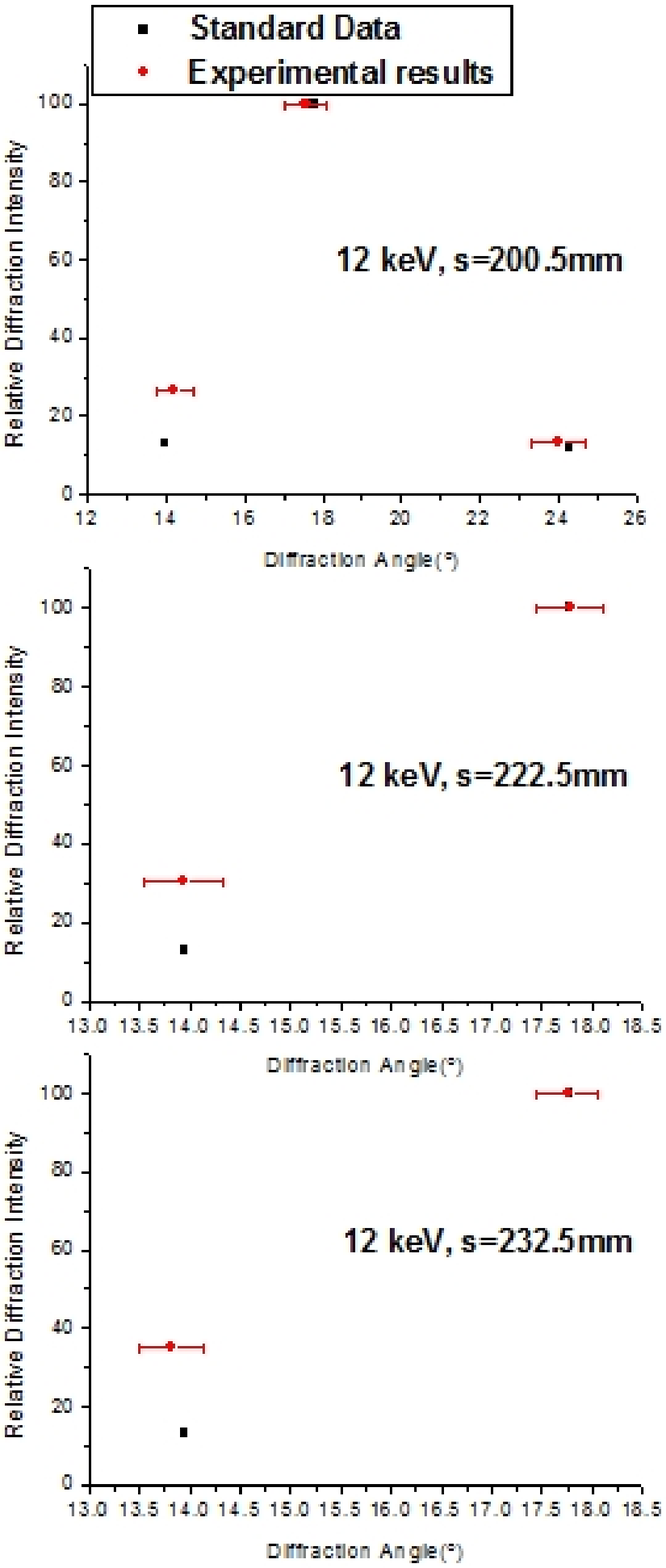}
\figcaption{\label{distance} Diffraction Angle of sample at different distances(Red points are experimental 2$\theta$, Black squares are pdf-2 standard data).}
\end{center}

When $S$=400 mm, it can be deduced that $\sigma_{2\theta}^r/\sigma_{2\theta}>50\%$. Based on the expected relative $\sigma_{2\theta}^r$ value, it makes corresponding $\sigma_{2\theta}<1\%$, which is the expectation of further measurement.

In spite of the fact that diffraction angles acquire higher precision with longer $S$, the beam intensity would limit the longest distance. For 8 keV X-ray, the mass attenuation coefficient for air is 9.9 cm$^2$/g(18 $^\circ$C, \cite{lab9}). It can be deduced that the intensity of X-ray would be attenuated to $\sim50\%$ at the distance of 600 mm, which defines the limit. In conclusion, to improve the precision of measurement without damaging detection efficiency, a optimal scheme proposed is to keep the distance $S$ between 400-600 mm.

\end{multicols}

\begin{center}
\tabcaption{ \label{mark}  Representation of relative $\sigma_{2\theta}$ and relative $\sigma_{2\theta}^r$ on experiment and calculation.}
\vspace{-3mm}
\footnotesize
\begin{tabular*}{170mm}{@{\extracolsep{\fill}}cccc}\toprule
$S$/mm & experimental relative $\sigma_{2\theta}$ & calculated relative $\sigma_{2\theta}^r$ & $\sigma_{2\theta}^r/\sigma_{2\theta}$  \\
\hline
200.5 & 3.27\% & 1.11\% & 33.8\% \\
222.5 & 2.57\% & 1.00\% & 38.8\% \\
232.5 & 2.20\% & 0.95\% & 43.3\% \\
300   &        & 0.74\% & \\
400   &        & 0.55\% & \\
500   &        & 0.44\% & \\
600   &        & 0.37\% & \\
700   &        & 0.32\% & \\ \bottomrule
\end{tabular*}%
\end{center}

\begin{multicols}{2}
\section{Conclusion}
A one-dimensional position sensitive detector has been developed with 200 readout strips and the position resolution of 138 $\mu$m obtained. The beam test of the detector has been accomplished on synchrotron radiation source. Analyzed the data of X-ray diffraction, The diffraction angles in the experiment were in accordance with standard data and the less than 1\% of relative errors between experimental results and standard data were obtained. Based on the beam test results, a optimized solution of the distance between the sample and detector were given. That is confining the distance between the sample and detector to 400-600 mm.

The experiment indicates that single chamber could be used on the measurements of X-ray diffraction on SR source , and 1D single wire chamber has good precision but simple structure and large measuring range. The further test will measure diffraction angles with improved scheme, while changing the kind of working gas, as well as the kind of powder sample, such as ZnO.

\begin{center}
\includegraphics[width=8.5cm]{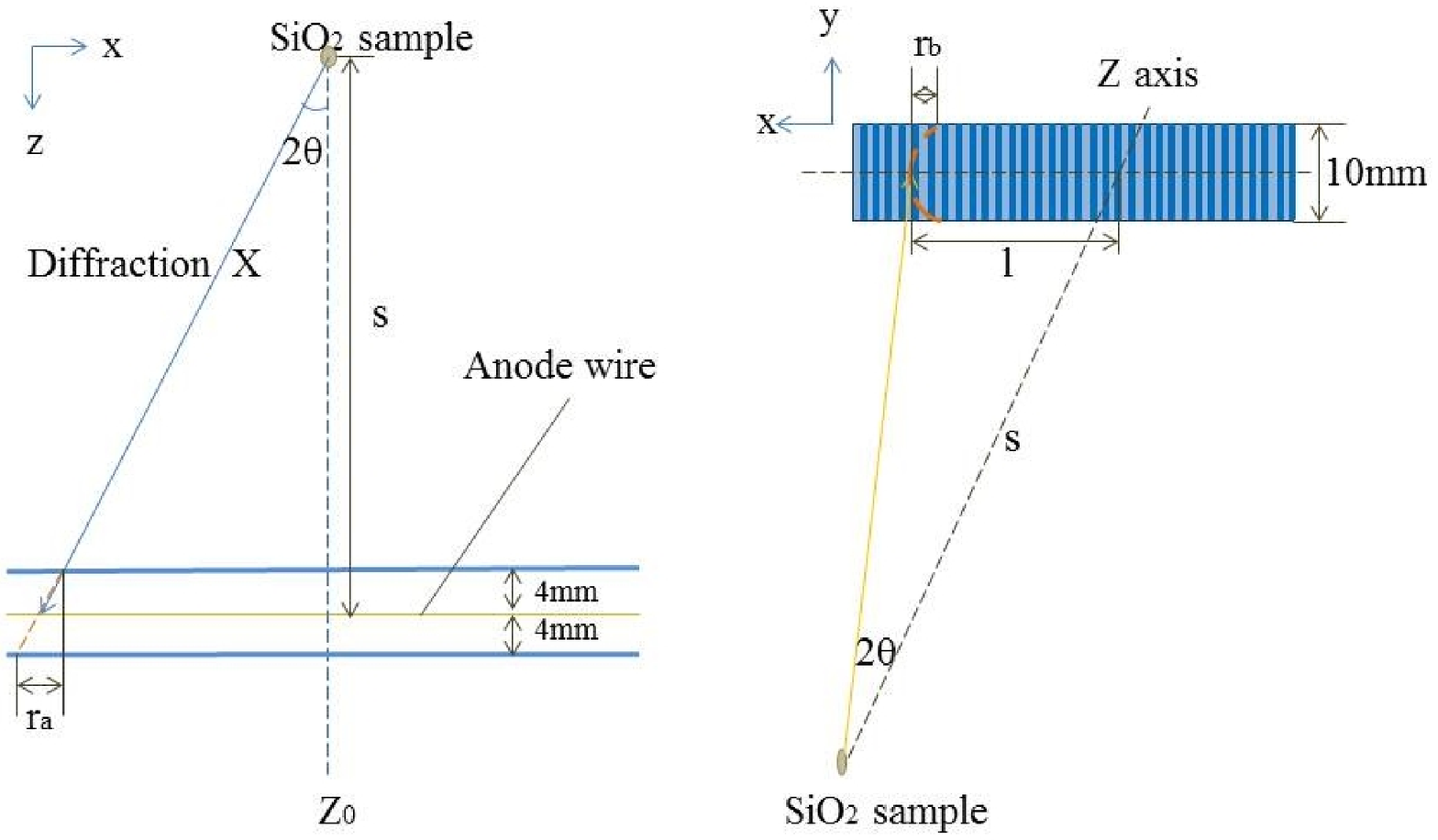}
\figcaption{\label{error}  (a)The plain receiving surface of detector produced maximum position error $r_a$ on x axis; (b)The 30 mm length of readout strips on y axis produced maximum position error $r_b$.}
\end{center}

\section{Acknowledgements}
We gratefully acknowledge the contributions of all colleagues at 1W2B laboratory of Beijing Synchrotron Radiation Source. And we thank Wang Ya-jie prepared samples for the experiment.
\end{multicols}

\vspace{10mm}

\begin{multicols}{2}

\end{multicols}

\clearpage

\end{document}